# Powering and Structuring the Universe Starting at Combination Time* - On the Electrodynamics of the Big Bang Universe

Howard D. Greyber
2475 St.    Lawrence   Drive,   San   Jose,   CA   95124
hgreyber@yahoo.com

Applying the known physics of plasma to the evolution of galaxies and quasars in the Early Universe, a unique "Strong" Magnetic Field Model (SMF) was created that explains the origin of a very large-scale primordial magnetic field in each Supercluster and the observed large-scale structure of galaxies. This physical model, involving both gravitation and cosmical magnetism, explains the existence of significant magnetic fields in galaxies. An intense highly relativistic gravitationally bound current loop (Storage Ring) is formed by gravitational collapse explaining the nature of the AGN/Quasar Central Engine, galactic structure and radio, optical and X-ray jets.

## 1. INTRODUCTION

Most astrophysics books give short shrift to the time, an age of about 400,000 years in the Big Bang Model of the evolution of the Universe, i.e. Combination Time, when the first atoms were formed. They simply say that at precisely that time the entire Universe changed from fully ionized plasma to become neutral gas in the Universe's long Dark Ages. Combination Time is much more complex than that (1.2.3).

The famous observed cosmic background radiation (CMB), predicted by Gamow, Alpher and Herman around 1950, and first observed about 1964, was emitted at Combination Time. Only, so the conventional story goes, by the effect of Gravity acting to pull matter onto very tiny density fluctuations that, allegedly, amazingly, survived, left over from far back in time, Inflation Time (a Big Bang age of 10exp-35 seconds), do the first stars, quasars and galaxies form hundreds of million years later when the Dark Ages ended.

The dubious ability of pure Gravity alone, to produce huge $10^9$ solar masses quasars within 700 million years, massive stars within 200 million years and significant early large-scale structure observed by Steidel et al, already at $Z = 3$, (4), is not discussed in today's common model. Nor is the fundamental problem of the origin of magnetic fields in the Universe touched upon, despite the prescient comments of Eugene Parker (5) about the impossibility of explaining the formation of observed large-scale magnetic fields in the Universe by present-day physics. The facts of plasma physics are not really applied to the origin and evolution of the large-scale structure of galaxies, and of large-scale magnetic fields in our Universe. (However most books do mention that our Universe is today made up of over 99% plasma)

Also not explained are the carefully reasoned conclusions of Jaan Einasto and nine other scientists (6): "We present evidence for a quasi-regular three dimensional network of superclusters and voids, with the regions of high density separated by 120 Mpc. If this describes the distribution of all matter (luminous and dark), then there must exist some hitherto unknown process that produces regular structure on large scales." In my opinion, my "Strong" Magnetic Field model (SMF) is their "hitherto unknown process". De Lapperent (7) describes the large-scale galaxy distribution as "sharp sheet-like structure delineating voids" Gregory (8) concluded "The large-scale structures we observe today had to be impressed somehow onto the proto-galactic material before it collected into galaxies". One model, SMF, explains all this evidence as well another puzzle, the origin of magnetic fields in our Universe.

The belief that tiny fluctuations produced at Inflation Time ($10^{-35}$ seconds) remain unchanged, after basic phase transitions, like separating the weak force from the EMF, after quarks coalesce into nucleons, and after nucleons form into nuclei, seems extremely unlikely to me. By contrast, in SMF the fluctuations are created and grow right at Combination Time, a Big Bang age of about 400,000 years. Inflation, if true, can be justified by its explanation of the horizon and flatness problems, and the total lack of observed magnetic monopoles, without, supposedly solely by Gravity, also "explaining" the rapid formation of stars, quasars and galaxies.

In my SMF model, the word "Primordial" refers to processes occurring at and after Combination Time, like George Gamow used it. The word "Strong" has been used by me in SMF since 1961 in precisely the same sense that the Russian physicist Y. Zel'dovich used it in 1983, "A major challenge is to understand strong magnetic fields whose energies greatly exceed those of hydrodynamics motions"(9). This means that the "Strong" magnetic field could, sometimes, appear extremely weak by Earth-like standards.

This interplay of Gravitation and Magnetic Field explains in SMF the wide variety of the dynamics and topology of objects of galactic dimension, as was emphasized by me in 1964 to 1967 and 1988-2001. Let us remember that the electromotive force (EMF) is $10^{40}$ times stronger than Gravity, and also that EMF is the ONLY other long-range force known in physics. This means that, in a Plasma, magnetic fields produced by even very tiny generated currents will locally dominate the motion of the plasma.

It is crucial to observe that even after Combina-tion Time the Universe is still a Plasma according to Lyman Spitzer's definition. The Debye shielding distance, h, in centimeters, is given by

$$h = 6.90(T/n)^{0.5}$$

where T is the temperature in degrees Kelvin, and n is the number of electrons per cubic centimeter. Spitzer states that "If $h$ is small compared to other lengths of interest, an ionized gas is called a Plasma" (10).

According to P.J.E. Peebles, those patches of the Universe where almost all the ions and electrons have combined still have a residual ionization of 0.01 to 0.0001 under various assumptions (11). Just after Combination Time, T would be about $10^4$ and n would be about $10^{-10}$. This yields h around $10^8$ cm. clearly more than ten orders of magnitude less than any other lengths of interest in galaxy formation.

Thus we conclude that the Early Universe was always Plasma, even in its Dark Ages. That gigantic electric currents and strong long-range magnetic fields can exist in the Universe for times long compared to the age of the Universe, was pointed out long ago by the pioneering Swedish astrophysicist, Hannes Alfven. Other theoretical models for the Early Universe, other than SMF, basically rely on only the gravitational field with various different strange ad hoc ideas added on.

## 2. GENERATION OF CURRENT AND PRIMORDIAL LARGE-SCALE MAGNETIC FIELDS IN THE SMF MODEL FOR THE EARLY UNIVERSE

Greyber (1967a, 1989) analyzed the effect of dissipation over time on a bounded turbulent medium. He showed that, over long times, small wavelengths dissipate their energy faster than large wavelengths. Thus about a million years after Combination Time, one is left with only huge plasma volumes pushing against other plasma volumes.

However, a better approach is to examine the phase transition at Combination Time, i.e. the relatively sudden change at Combination Time in the Big Bang Model, from fully ionized plasma to an almost neutral plasma, releasing energy in the form of a flood of photons. In a long article on the "Dynamics of First Order Transitions", Gunton (12) states that the problem of pattern formation in systems which are spontaneously attempting to reach equilibrium provides a fascinating example of nonlinear, nonequilibrium phenomena. In the theory of first order phase transitions there are two different types of instability. One is nucleation, the slow movement of matter towards a positive density fluctuation.

For our purpose we consider the other one, an instability against infinitesmal amplitude, nonlocalized (long wavelength) fluctuations which leads to the rapid initial decay of an unstable state. This instability is called SPINODAL DECOMPOSITION, named by American physicist J. Willard Gibbs in 1906. The first successful theory of Spinodal Decompositioin was created by Hillert, and later developed by John W. Cahn at N. I. S. T. Cahn developed a linearized deterministic theory that predicts an exponential growth of the fluctuations, rather than the much slower growth of nucleation. (Cahn won the National Medal of Science for this research) This exponential growth determines the morphology of the Universe at Combination Time.

Clearly this fast growth progresses only until one border encounters the border of another plasma volume. It is unphysical to assume that the whole infinite Universe changed phase at precisely the same time. The real process is similar to that in a quiet lake as the temperature falls extremely slowly through zero degrees Centigrade. Some patches of the lake will supercool, staying liquid, while other patches will freeze.

Similarly, at Combination Time, the Universe will consist partly of patches of only slightly ionized plasma, and partly patches of fully ionized plasma. In those patches that have become almost neutral, about 10eV per atom is released. The flow of photons will penetrate only a very short distance into a neighboring fully ionized patch because the photon mean free path in fully ionized plasma is far shorter than in almost neutral plasma. The photons heat this border layer up, compressing the patch of fully ionized plasma, thus increasing dramatically the seed magnetic field inside, and thus dramatically increasing the jump in magnetic field across the boundary. For a discussion of seed magnetic fields, see Sir Martin Rees (13)

It is reasonable to assume slight differences in the seed magnetic fields and slight differences in the matter density across such interfaces. Of course, among the infinite number of patch interfaces at this time in our infinite Universe, there will be an infinite subset of interfaces where a hotter denser plasma with a stronger magnetic field presses against a less dense, cooler plasma volume with a weaker magnetic field. These particular spatially curved boundaries are Rayleigh-Taylor Stable.

Let us recall that it follows from Maxwell's Equations that there must be an electric current at the boundary between two adjacent volumes of plasma with different magnetic field intensities inside. The electric current is proportional to the jump across the boundary of the component of the magnetic field parallel to the boundary on each side. Enormous electric current amplification continues at these stable boundaries.

It is important to note that there are different equations of state for magnetic field and for matter. While magnetic pressure varies as B squared, matter pressure goes as the density to some power like 1, 4/3, or 5/3 For example, assuming an isothermal plasma, a twenty-fold doubling, i.e. a compression of $10^6$ is possible over a few million years. This process increases v in the expression for current j = nev. Actually, due to the three dimensional wrinkling of sheet-void boundaries, current amplification in certain local volumes (in corners) will be very much larger.

However, simultaneously, the number of current carriers, n , will also increase dramatically due to (a) the well-known Pinch Effect that draws in charged particles until they join the current flow at the boundary, and (b) random particles which cross the current flow and are captured. If n has a twenty-fold doubling along with that of the density, then an increase of current, and of magnetic field, of $10^{12}$. is expected. Thus a seed magnetic field of $10^{-21}$ gauss could be amplified to about $10^{-9}$ gauss.

In addition, as the mass density at the current-carrying interfaces rises compared to the average density on either side, Gravity begins to act to draw both neutral and charged plasma into the interface region. Notice that even if the SMF process has only increased the matter density at the spatially curved boundaries by only 0.00001 of the average matter density in a patch, that is sufficient to imprint the observed structure on the proto-galactic matter. This is because Gravity continues to attract matter into the relatively thin sheets of matter around voids long after the pressure differences are gone.

Eventually the density along the sheets of matter rises until at some local volumes, critical density for gravitational collapse of a plasma cloud that will form a galaxy or a star is reached. As I pointed out in the 1964 I.A.U., shock waves from the initial gravitational collapse will envelope nearby matter concentrations that are very close to critical density, inducing them to collapse, thus favoring the formation of clusters of galaxies and clusters of stars.

Thus quasars and galaxies are formed along spatially curved thin sheets around huge voids, matching the astronomical observations of large-scale structure. Since the Pinch Effect was proven to be unstable by Kruskal and Scharzschild (14), the large majority of the matter will remain anchored, as DARK MATTER, near the DeVaucouleurs Superclusters in some form.

## 3. GENERATION OF GRAVITATIONALLY BOUND CIRCULAR CURRENT LOOPS, i.e. STORAGE RINGS, DEFINING THE CENTRAL ENGINE OF QUASARS AND GALAXIES

The Storage Ring concept was created by me based on my growing conviction of the importance of dipole magnetic fields in galaxies, on the unpublished 1963 circuit theory derivation by Menzel and Greyber, and on the pioneering MHD analysis by Mestel and Strittmatter (15). The great usefulness of this Storage Ring concept in explaining astronomical observations was discussed by me at length in several articles, Greyber (16-23), including the arguments for generation, equilibrium and stability of the current loop. In (20) and (21), I present a Table comparing the predictions of the SMF Storage Ring model with those of the conventional rotating accretion disk model involving only Gravity. Clearly, the SMF model fits the astronomical observations overwhelmingly better.

As magnetofluiddynamics dictates (24,25), the intense current loop maintains a cavity around itself inside the massive, slender toroid of plasma, which is bound to the current loop by the Maxwell "frozen-in" magnetic field condition. The bursting force of this very strong unified magnetic field system is in equilibrium with the strong gravitational force between the massive slender toroidal plasma and the central massive object, probably a black hole.

Thus Nature, or *der leibe Gott*, in this novel paradigm, taps a small fraction of the gigantic energy of gravitational collapse of an entire pre-galactic/pre-Quasar plasma cloud, storing it in a Storage Ring, which defines the galactic/quasar central engine. It is this SMF Central Engine (see the Figures) which produces jets and galactic spiral arms, and the phenomena of quasars, galaxies and gamma ray bursts.

Galaxies accrete matter from neighboring smaller galaxies, and from Oort clouds. Most plasma accretes down the dipole magnetic field, against the adiabatic invariant, into the galactic nuclear region. There, in a spiral galaxy, where the ratio of magnetic energy to rotational energy is low, the matter in the nuclear region is expelled into usually two spiral arms, forming dust clouds and then bright O and B stars. This matches the observations by Walter Baade of the nuclear region of M31. Note that for a very low ratio, the bulk diamagnetic plasma may tend to simulate a disk. However in an active galaxy, where the ratio of magnetic energy to rotational energy is quite high, like a radio galaxy or quasar, the matter is expelled in successive blobs along the rotation axis, forming the observed jets.

The subtle arguments for Stability of the Storage Ring are worth expanding on. Macroscopically, there are two important hints. Both thick and thin rotating accretion disk structures exhibit many serious instabilities. Using hydrodynamics, O. M. Blaes (26) analyzed constant angular momentum tori , demonstrating that, for extremely slender tori (such as the SMF storage ring) any instability, becomes vibrational, i.e. Stable. Probably the same result will occur for the MHD case. More evidence is adduced by the fact that the two dimensional current loop *with the minimum perimeter*, i.e. a circle, proves to be the most stable.

However, the microscopic picture seems the most compelling. Due to the coherence, one part of an undisturbed, perfectly steady loop current will not radiate in the magnetic field of another part. However, if a fluctuation, or "bump", appears somewhere along the loop, the particles in the "bump", now out of coherence, will suddenly radiate furiously in the extremely strong magnetic field, the energy in the fluctuation will dissipate rapidly, and the ring will return quickly to a circular loop. Thus the intense, completely coherent, highly relativistic current loop (Storage Ring) is uniquely stable. Please note that the only potential causes of a "bump", a star or a planet, are tiny compared to the huge loop perimeter.

A current loop in the extremely high vacuum of interstellar space will begin with *Jitter Energy* described by a temperature perpendicular to the current direction, Tperp, and a temperature parallel to the current direction, Tpar. However relatively quickly, the Tpar. energy will be transferred into Tperp. energy, which in turn will diffuse to the surface of the current loop and radiate away into space as cyclotron radiation. Beam physics thus leaves an undisturbed, completely coherent current loop, i.e. "gravitationally bound current loop" or "Storage Ring", as an unusual unique object, speeding at extremely relativistic speeds, but with a remarkable, crystal-like, structure, in the central engine of galaxies and quasars. If left undisturbed, the only radiation loss from the loop is the extremely small curvature radiation, previously calculated from J. Schwinger (27, and Tables in 20,21).

## 4. DISCUSSION AND CONCLUSIONS

Greyber's SMF Model provides astrophysics with a New Classification of Galaxies besides that created by Edwin Hubble, i.e. the morphology and energetics of objects of galactic dimension are classified and determined by the Ratio of magnetic energy to rotational energy in the particular object (18,23). This Ratio is highest for quasars, BL Lacs and blazars, decreasing steadily for giant elliptical radio galaxies, then less for Seyfert and Markarian galaxies, is low for Spiral galaxies, and even lower for ordinary elliptical galaxies. However the activity observed is also a function of the matter accretion rate at the time observed.

One outstanding puzzle of cosmology is *"When and How did Magnetic Fields Originate?"*. Two relevant research results have occurred recently. One is the discovery that there exist "large-scale magnetized regions of space *outside* clusters of galaxies" (28). Thus the real question, as (28) poses it, is whether this magnetic field is *primordial*. (28) answers No, claiming that young dwarf galaxies seed the intergalactic medium with magnetic fields. (28) assumes very strong dynamo amplification will occur at the highest redshifts, although the galactic dynamo model has serious defects.

Physicist Walter Elsasser in 1934 related to a friend that Thomas G. Cowling was attempting to make a theory for an axially symmetric astrophysical dynamo. "If that simple idea does not work", remarked Albert Einstein, "then dynamo theory will not work". The result of the research was negative, yielding Cowling's famous antidynamo theorem. Agreeing with Einstein, I argue that very large-scale primordial magnetic fields were created in each Supercluster far earlier, at and after Combination Time by the SMF processes, not by a "dynamo" process.

The second relevant result was deduced from the observations of D. Hutsemekers and H. Lamy (29) titled "Confirmation of the existence of coherent orientations of quasar polarization vectors on cosmological scales". They found "the (electric) polarization vectors in region A1 are apparently parallel to the plane of the Local Supercluster". They point out that a distinct possibility is that the quasar structural axes themselves are coherently oriented.

(29) concludes that "the presence of coherent orientations at such large scales seems to indicate the existence of a new, interesting effect of cosmological importance". Obviously random flows surging from individual dwarf galaxies could not possibly explain coherent orientations over a large fraction of the size of a Supercluster. The only clear explanation left is that those quasars observed were formed in the presence of a relatively strong, very large-scale primordial magnetic field formed at about the same time, in the very same processes that imprinted the large-scale observed structure of galaxies in the Universe, i.e. at and after Combination Time.

The SMF central engine clearly predicts that jets emerge in successive blobs, as K. Kellerman and P. J. Wiita believe. Kellerman concluded "typical morphology shows a compact core with one or more blobs moving away in a single direction"(30). Accretion of bulk plasma inevitably forces open either one of the two throats and a blob of plasma is ejected along the rotation axis, producing two-sided jets, (see the Figures) But if the accretion is extraordinarily large, one throat cannot close and one-sided jets occur for that radio source, agreeing with observations. However some astronomers claim that blobs are merely the result of shock waves. This SMF prediction could use more confirmation.

Another specific prediction of the SMF Model is that the Spiral Arm magnetic field is in one direction above the plane of the galaxy, and in the opposite direction below, with a neutral sheet along the plane of the galaxy. The same configuration is found in the Earth's Magnetotail, which the IMP satellite found is quite stable. Prof. S. Chandrasekhar kindly told me that the uniform spiral arm magnetic field model assumed by him and Fermi, was chosen simply for simplicity, *not* due to physical intuition.

Long ago, Lyman Spitzer was concerned that mas-

sive stars might not be able to be formed in the presence of spiral arm magnetic fields of $10^{-6}$ gauss. In our SMF Model they form easily when formed in the low magnetic field close to the galactic plane. Using Faraday rotation, D. Morris and G. L. Berge (31) deduced that in our spiral arm, where our Sun is located, the magnetic field tends to be roughly in opposite directions above and below the galactic plane. More confirmation is needed for this prediction of SMF.

Observations have shown that narrow, very strong, straight jets are emitted by some newly forming stars. The increasing dipole magnetic field created temporarily by a storage ring formed in the newly formed star will produce and confine a narrow, very long, straight jet (32). However when the density inevitably becomes too high, either the ring is destroyed, or the current-carrying plasma ring is buried inside the newly formed star, becoming the source of primordial stellar magnetism in that star, perhaps in our Sun.

Since 1961, SMF has emphasized repeatedly that it is not necessary to assume Equipartition in the central engine of quasars and galaxies. As Prof. Charles H. Townes puts it (personal communication), "Equipartition exists only when the physics demands it does". Clues to the importance of strong magnetic fields in quasars can be found in early research by G. Burbidge and F. Hoyle (33). They pointed out, for instance, that for the synchrotron process to be dominant, "this means that quasi-stellar objects must contain - millions of separate points where electrons are injected into a strong magnetic field". SMF's slender Storage Ring obviously meets this criterion.

We note that SMF discovered two novel concepts used in Nature for production of electric current, both different from the Faraday/Henry discovery that powers our electric civilization, and different than the minor processes listed in the Lars Onsager Matrix found in Thermodynamics. The first uses the energy of the series of photon flows in the Big Bang Model at Combination Time, acting on remaining patches of fully ionized plasma. The second uses the energy of gravitational collapse of a pre-galactic or pre-quasar plasma cloud permeated by a uniform primordial magnetic field to create a Storage Ring.

It is pertinent to observe that the SMF Model agrees fully with the comment by Sir Martin Rees that "the phenomena of quasars and radio galaxies cannot be understood until they are placed in the general context of galactic evolution" (34). By an original, unique "Strong" Magnetic Field Model (SMF), invoking the known physics of plasmas, several extremely important puzzles in Astrophysics can be explained.

## Acknowledgements

It is a pleasure to thank the late Donald H. Menzel for friendly sage advice and warm encouragement, John A. Wheeler for insightful comments, and Gart Westerhout for permission to use the U. S. Naval Observatory Library.

## APPENDIX A

How likely is it that the Inflation fluctuations at an age of $10^{-35}$ seconds remain scale-invariant over 50 powers of ten growth in scale, despite several phase transitions (like separating the EMF force from the weak force), quarks assembling into nucleons, nucleons into nuclei, and by Combination into atoms, compared to the fluctuations produced by the SMF processes which have to stay scale-invariant for only about 9 powers of ten of growth in scale, while all is quiet during the Dark Ages, until the first gravitational collapses produce light. To this physicist, the latter is far more likely.

In SMF, the EMF force (by the Pinch Effect) first pulls the 3D ionized particles into spatially curved sheets. Then Gravity gradually takes over, attracting both charged and neutral matter. Thus the EMF force helps Gravity achieve the first stars by only 220 Million years (estimated by WMAP). Recall that E. S. Phinney (CalTech) was reported in Sky & Telescope to be concerned that with CDM one would not have time enough to form huge $10^9$ solar mass quasars in only one billion years, as observed. Now, from Sloan, there are a few more huge quasars, including one at an age of only 700 Million years. Can Gravity alone, a very, very weak force, explain those observations?

Note also that in my second diagram of the Central Engine, as John A. Wheeler calls it, when the key ratio is low, the bulk plasma simulates more and more a disk confined on the horizontal axis by magnetic field. It was first proven in 1952 (when I was at Princeton University) by a pioneering computer calculation, MacBeth, created by Prof. Wheeler that I helped run, that in Rayleigh-Taylor Instability, LONG wavelengths dominate in the Nonlinear range, contrary to what happens in the linear range. Clearly the longest wavelength perturbation for a disk is to become an oval. Thus SMF predicts that in most spiral galaxies, two arms are formed, not one or three, just as observed in Nature. The spiral arm magnetic field produced reverses across the galactic plane, a stable configuration, as seen in the Earth's Magneto-Tail repeatedly by the IMP satellite.

In SMF, th logical place for fluctuations to be produced are by the Unstable Pinch Effect, (see Kruskal & Schwarzschild, Ref. 14) acting due to the increasing current in the boundary between the fully ionized plasma (FIP) and the slightly ionized plasma (SIP) at COMBINATION TIME. Los Alamos has done many Pinch Effect experiments for the Controlled Thermonuclear Research (CTR) program, and one can see in their data the fluctuations produced.

## APPENDIX B

It is interesting and worth noting that, in the SMF model, we have said nothing about what happens after Spinodal Decomposition to the supercooled FIP (fully ionized plasma patch) which remains a sort of false vacuum, far out of equilibrium, perhaps representing a negative pressure. The first SMF process generates current and pulls in charged particles to the boundary. But it acts only along the boundaries of the FIP patch, as the series of photon flows compress the FIP. The major flow, of course, occurs when adding an electron to make a neutral hydrogen atom.

If the typical Supercluster were 360 Million light years in length today (120Mpc), then back at Combination Time, the FIP patch then was roughly 10,000 to 15,000 light years long. It appears possible that some (or all) of these FIP patches remain FIP and supercooled still today, located in some voids, representing today a cosmic repulsion. Gradually as the expansion slows due to Gravity from that at the origin of the Universe, then at some point this cosmic repulsion would overcome the effect of Gravity on the expansion of the Universe, and the Universe would begin an accelerating expansion. According to the remarkable observations by two groups, one headed by Saul Perlmutter and the other by Robert Kirshner, this transition occurred at about an age of nine billion years in the Big Bang Model.

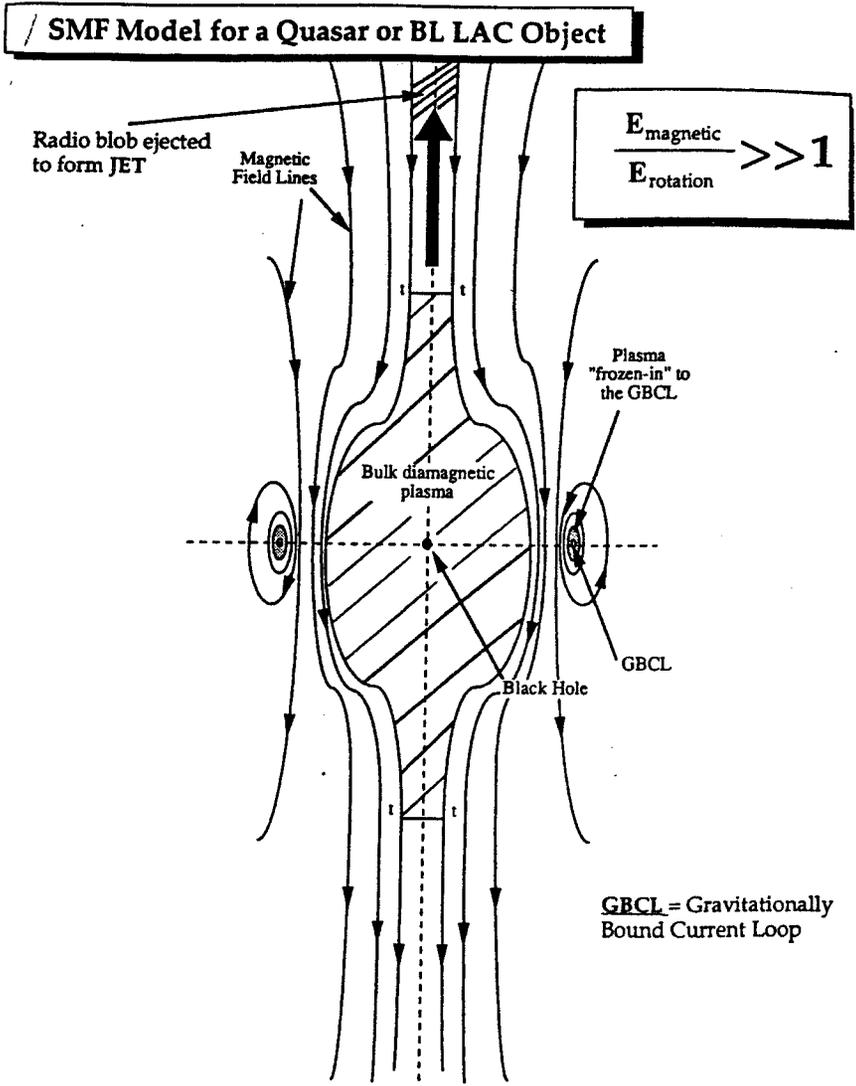

The Strong Magnetic Field AGN-*Central Engine* Quasar-Galaxy Formation Paradigm
H. D. Greyber

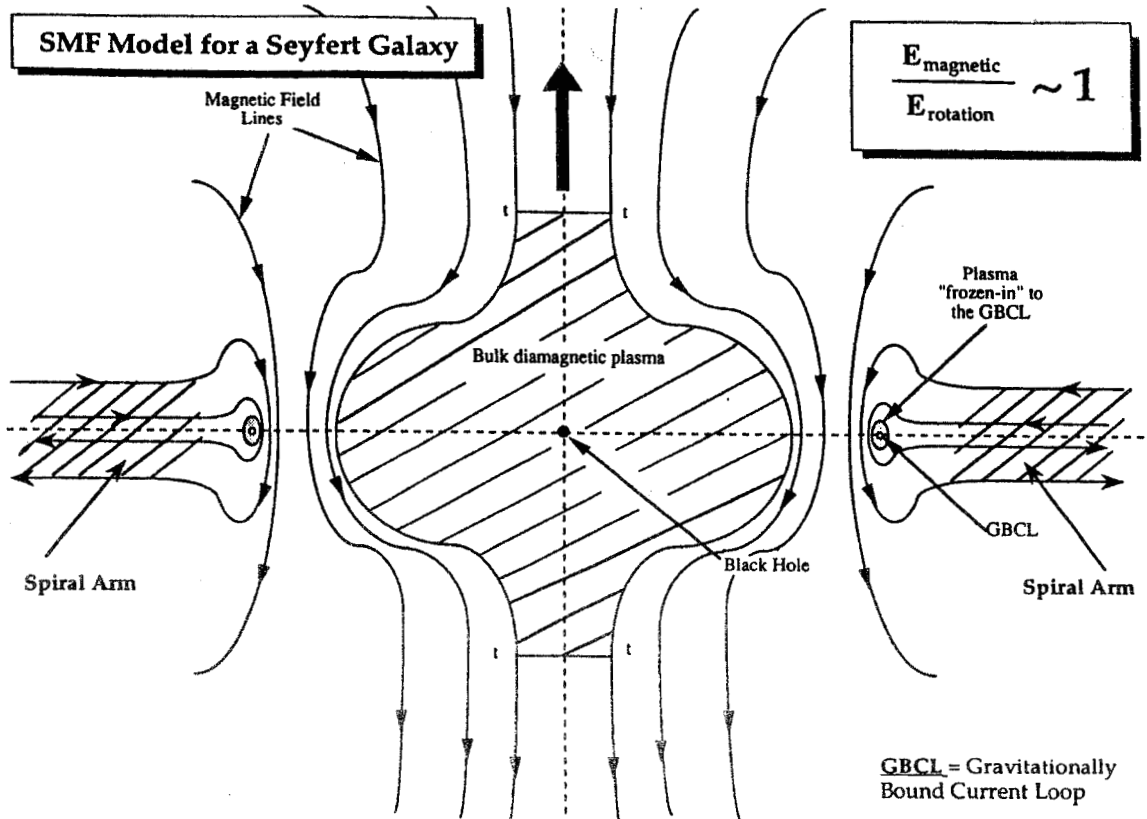

The Strong Magnetic Field AGN-*Central Engine* Quasar-Galaxy Formation Paradigm
H. D. Greyber